\documentclass[twocolumn,showpacs,amsmath,amssymb]{revtex4}

 \usepackage{graphicx}

\usepackage{amssymb}
\newcommand{\be}{\begin{equation}}
\newcommand{\ee}{\end{equation}}
\newcommand{\bfig}{\begin{figure}}
\newcommand{\efig}{\end{figure}}

\begin{document}


%

\title{Magnetization, Nernst effect and vorticity in the cuprates}

\author{Lu Li$^1$, Yayu Wang$^1$, M. J. Naughton$^2$, Seiki Komiya$^3$, 
Shimpei Ono$^3$, Yoichi Ando$^3$, and N. P. Ong$^1$\footnote{Symposium, Int. Conf. Magnetism, 
Kyoto 2006, J. Magn. Magn. Mater., \emph{in press}}
}
\affiliation{
$^1$Department of Physics, Princeton University, Princeton, New Jersey 08544, U.S.A.\\
$^2$Department of Physics, Boston College, Chestnut Hill, Massachusetts 02467, U.S.A.\\
$^3$Central Research Institute of Electric Power Industry, Komae, Tokyo 201-8511, Japan
}

\date{\today}      
%


\begin{abstract}
Nernst and magnetization experiments reveal the existence of a large region 
of the cuprate phase diagram above the $T_c$ curve in which vorticity and weak 
diamagnetism exist without phase coherence.  
We discuss the implication that the transition at $T_c$ is caused by the loss of 
long-range phase coherence caused by spontaneous vortex creation.   
Below $T_c$, these measurements provide an estimate of the depairing field $H_{c2}$ 
which is found to be very large (40-100 T depending on doping).  
We discuss the high-field Nernst and magnetization results, binding energy, and the phase
diagram of hole-doped cuprates.  Some new magnetization results on the vortex liquid 
in very underdoped LSCO in the limit $T\rightarrow 0$ are reported as well.
\end{abstract}

\maketitle                   


%
%
%
\section{Introduction}\label{intro}
A striking characteristic of a superconductor is the stiffness of the macroscopic
wave function $|\Psi|\exp{i\theta}$ against distortions of its phase $\theta$
(London rigidity~\cite{Schrieffer,Anderson66}).
The long-range phase coherence sustained by the phase stiffness
-- analogous to the shear rigidity of an ordinary solid --
is responsible for superfluid properties such as the Meissner effect.
The loss of shear rigidity turns a solid into a liquid.  What is the analogous
transition in a superconductor?  
In BCS (Bardeen-Cooper-Schrieffer) superconductors, where the pair amplitude $|\Psi|$ vanishes at
the critical temperature $T_c$ (in zero field), this question does not arise because 
phase coherence is maintained right up to $T_c$, above which 
the condensate (the ``stuff'') disappears.

However, in the Kosterlitz-Thouless (KT) transition in two-dimensional (2D) systems, loss of phase 
coherence occurs at a temperature $T_{KT}$ lower than 
the temperature at which $|\Psi|$ vanishes~\cite{KT}.  The phase-coherence collapse results
from the spontaneous unbinding of vortex-antivortex pairs driven by entropy gain.
The 2D superconductor becomes unstable to the spontaneous appearance 
of \emph{mobile} vortices at $T_{KT}$.  Above 
$T_{KT}$, $|\Psi({\bf r})|$ remains finite but the rapid diffusion of (anti)vortices leads
to strong (singular) fluctuations in $\theta({\bf r})$.  The condensate, with its phase
rigidity restricted to the short length scale $\xi_{\phi}$, corresponds to the ``liquid'' 
in the above analogy ($\xi_{\phi}$ is the phase correlation length).  More generally, 3D superconductors
with highly anisotropic coupling, low superfluid density $\rho_s$ and large pair-binding
energy may suffer a similar loss of phase coherence by vortex(loop) creation.  
We will call such transitions the phase-disordering scenario.  

From the start, the assumption that the cuprates follow the BCS scenario has been deeply 
entrenched and surprisingly difficult to dislodge, given the absence of solid evidence.  
Nonetheless, a slim thread of evidence for phase disordering has always been present.
Very early predictions~\cite{Baskaran,Inui} that
phase disordering is dominant in the UD (underdoped) region could not be reliably tested 
(especially by high-field magnetization), and interest shifted to other issues.
The early muon spin relaxation ($\mu$SR) experiments~\cite{Uemura}, which showed 
that $T_c$ increases linearly with $\rho_s$ 
in the UD region may be seen, in hindsight, to be consistent with the phase-disordering scenario.  
In an influential paper, Emery and Kivelson~\cite{Emery} showed that cuprates differ from low-$T_c$
superconductors in having an anomalously low phase-disordering temperature.
The observation by Corson \emph{et al.}~\cite{Corson} of kinetic inductance up to 25 K 
above $T_c$ in ultrathin films of Bi 2212 re-focussed attention on this issue.  
It was unclear, however, if high-quality crystals behave differently.

Here we describe 2 experiments which have been highly effective in 
addressing the role of vorticity and phase-disordering 
in the phase diagram of the cuprates, namely the 
Nernst effect~\cite{Xu,WangPRB01,WangPRL02,WangSci03,WangPRB06} and torque 
magnetometry~\cite{WangPRL05,LiEPL05}.  
These 2 techniques probe directly the superfluid response, much like the $\mu$SR 
and kinetic inductance experiments.  However, they seem to be tailor-made for
tiny crystals.  Moreover, their resolution remains high even in fields 
up to 45 Tesla.  This has allowed significant progress in the task 
of measuring the magnetization curve
and determining the depairing field $H_{c2}$ in the hole-doped cuprates, which we
discuss at length.  We also describe recent results which 
probe the vortex liquid at low temperature $T$ = 0.5 K
as $x$ decreases below the critical value $x_c$.  

Throughout, we write Bi 2212, Bi 2201
and LSCO for $\rm Bi_2Sr_2CaCu_2O_8$, $\rm Bi_2Sr_{2-y}La_yCuO_6$ 
and $\rm La_{2-x}Sr_xCuO_4$, respectively.  UD, OP and OD stand for underdoped,
optimally doped and overdoped, respectively, while SC stands for superconducting.

\section{Vortex-Nernst effect}\label{vortexNernst}
In the vortex-liquid state, an applied temperature
gradient $-\nabla T||{\bf \hat{x}}$ causes the vortices to flow 
towards the cooler end of the sample with velocity $\bf v||\hat{x}$ 
(the magnetic field $\bf H||\hat{z}$).  
As each vortex core crosses a line drawn $\bf ||\hat{y}$, the difference of
the phases $\theta_1$ and $\theta_2$ at the ends slips by 2$\pi$.
This translates into a weak transient voltage pulse~\cite{Anderson66}. 
Integration of all the voltage pulses from a large number of vortices leads to a 
steady-state electric field given by the 
Josephson relation $\bf E = B\times v$, which may be observed as a Nernst signal $e_N\equiv E_y/|\nabla T|$.
Here, ${\bf B}=\mu_0({\bf H +M})$ with $\bf M$ the magnetization and $\mu_0$ the vacuum permeability.

\begin{figure}
\includegraphics[scale =0.47]{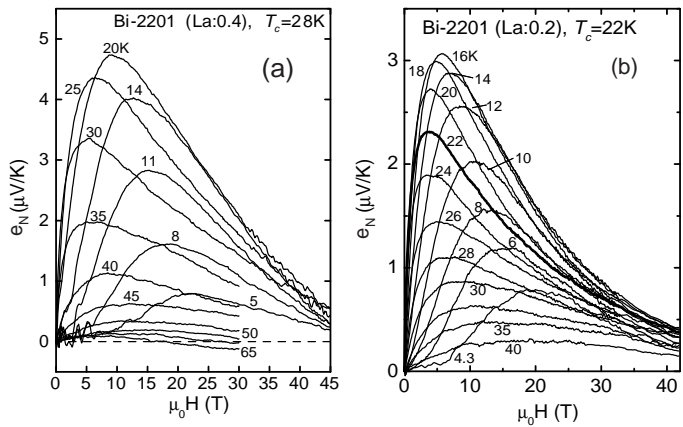}
\caption{  	\label{Nernst}
The Nernst signal $e_N$ vs. $H$ in OP Bi 2201 (Panel a, $T_c$ = 28 K) and in OD Bi 2201 (b, $T_c$ = 22 K). 
The characteristic hill profile of $e_N$ extends to $T$ high above $T_c$ (measured with $-\nabla T$ 
in the $ab$ plane and $\bf H||c$).  The depairing field $H_{c2}$ is estimated
by extrapolating $e_N\rightarrow 0$.  [Ref.~\cite{WangPRB06}].
}
\end{figure}

Figure \ref{Nernst} shows the Nernst profiles $e_N$ vs. $H$ in OP and 
OD Bi 2201 (Panels a and b, respectively)~\cite{WangPRB06}.  
The ``tilted-hill" profile which is apparent in intense fields is characteristic 
of all cuprates studied.  When $H$ exceeds the solid melting field $H_m(T)$, $e_N$ 
rises steeply to reach a maximum.   Above the peak, the condensate amplitude 
is progressively suppressed by $H$ until it decreases to a value approaching zero at the upper 
critical (depairing) field $H_{c2}$ ($\sim$48 T and $\sim$40 T in the OP
and OD samples, respectively).  Two noteworthy features are the smooth continuity of the signal 
across the zero-field $T_c$, and its persistence to temperatures high above $T_c$.
These 2 features are not observed in low-$T_c$ superconductors (or in the electron-doped
cuprate $\rm Nd_{2-x}Ce_xCuO_4$).  In the hole-doped cuprates, they 
provide strong evidence for the scenario that the transition at $T_c$ corresponds to 
the loss of long-range phase coherence rather than the vanishing of 
the amplitude $|\Psi({\bf r})|$.

\begin{figure}
\includegraphics[scale =0.5]{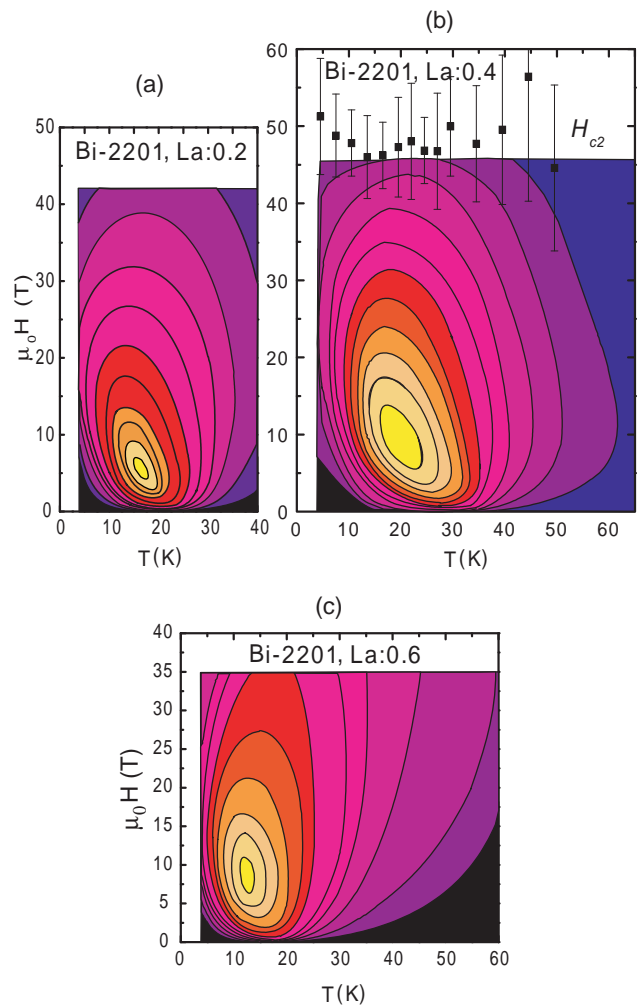}
\caption{  	\label{contour3}  (color online) Contour plots of the magnitude of the Nernst signal $e_N(T,H)$
in the $T$-$H$ plane for OD Bi 2201 ($T_c$ = 22 K, Panel a), OP Bi 2201 ($T_c$ = 28 K, Panel b) 
and UD Bi 2201 ($T_c$ = 12 K, Panel c).  
The magnitude of $e_N$ = 0 in the black regions, and increases in intensity through blue, violet,
red, orange to yellow (maximum intensity).  Note that the contour lines bulge to temperatures significantly
above $T_c$ at high fields.  In Panels (a) and (b), the high-field contours begin to form closed loops
implying the approach to $H_{c2}$.  Estimated values of $H_{c2}(T)$ are shown as solid squares
with error bars in Panel (b).  The axes in the 3 panels have the same scale.
}
\end{figure}

The contour plot of $e_N(T,H)$ in the $T$-$H$ plane provides an instructive way to view the vortex-Nernst signal.  
In Fig. \ref{contour3}, contour plots for OD, OP and UD Bi 2201 are displayed
in Panels (a), (b) and (c), respectively.   In each panel, $e_N$ increases from zero (black regions) through the blue,
violet and red regions to reach its maximum value in the yellow region.  
The vortex solid occupies the black wedge in the low-$T$, small-$H$ corner.  In Bi 2201, 
the vortex liquid, which dominates the entire $T$-$H$ plane, extends to very high fields.  
More significantly, it extends to temperatures considerably above 
$T_c$ in all 3 samples.  There is no evidence for an $H_{c2}$ line that goes to zero at $T_c$.
Instead, the inferred $H_{c2}$ is nearly $T$ independent [shown as solid squares in Panel (b)].
We return to the anomalous behavior of $H_{c2}$ below.

\begin{figure}
\includegraphics[scale =0.35]{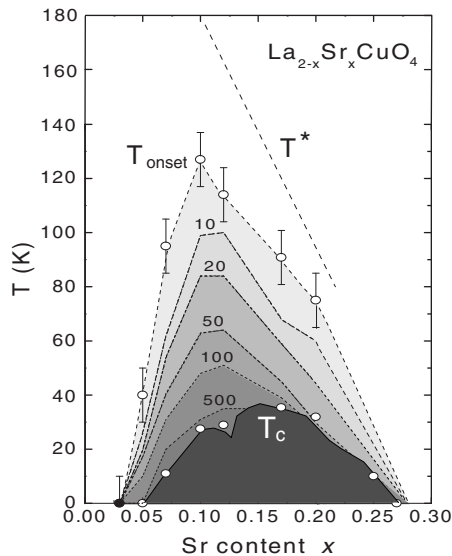}
\caption{  	\label{phasediag}
The phase diagram of LSCO showing $T_{onset}$ of the Nernst signal, the 
transition $T_c$ and the pseudogap temperature $T^*$.  In the ``Nernst" region between
$T_{onset}$ and $T_c$, vorticity is observed by the Nernst and torque magnetometry
experiments.  The numbers indicate $\nu = e_N/B$ in nV/KT 
(initial slope of the $e_N$-$H$ curve).  [Ref. \cite{WangPRB06}] 
}
\end{figure}
Above the SC dome in the phase diagram of LSCO, the vortex-Nernst signal is observed in 
the ``Nernst region" shown in gray scale in Fig. \ref{phasediag}.  The 
contour lines indicate the initial value of the Nernst coefficient $\nu = e_N/B$.
Clearly, the Nernst region is closely related to the SC dome defined by $T_c$ vs. $x$.
On the OD side, it terminates at $x\sim$ 0.25, while on the UD side, it reaches
to 0.03.  

The onset temperature $T_{onset}$ is defined as the temperature above
which $e_N$ cannot be resolved from the negative quasiparticle (qp) contribution \cite{WangPRB01}.  
As shown, $T_{onset}$ peaks at $x$ = 0.10 instead of the OP doping $x$ = 0.17
(all the contours also show this skewed profile so it is not due to difficulties in resolving 
$T_{onset}$).  The maximum value of $T_{onset}$ (130 K) is significantly lower than
values of the pseudogap temperature $T^*$ quoted for the UD region ($T^*$ is only roughly 
known in LSCO).

\section{Torque Magnetometry}\label{torquemagneto}
The vortex interpretation of the Nernst signals has received strong support 
from high-resolution torque magnetometry~\cite{WangPRL05,LiEPL05}.  
Because the supercurrent in cuprates is quasi-2D,
torque magnetometry is ideal for probing its diamagnetic response.  
If the angle $\phi_0$ between $\bf H$ and $\bf c$ is small,
the torque signal $\bf{\tau} = m\times B$ may be expressed as~\cite{WangPRL05,LiEPL05}
\begin{equation}
\tau = [\Delta\chi_p H_z + M(T,H_z)]VB_x,
\label{tau}
\end{equation}
where $V$ is the crystal volume, $\Delta \chi_p = \chi_z-\chi_x$ is the anisotropy of 
the paramagnetic (background) susceptibility and $M(T,H)$ the diamagnetic magnetization of interest
(we choose axes $\bf z||c$ and $\bf x$ in the $ab$ plane; hereafter we write $H_z=H$).

Above $\sim$4 K, we find experimentally that $\Delta\chi_p$ is dominated by
the paramagnetic van Vleck susceptibility $\chi_{orb}$.
Because $\chi_{orb}$ is $H$ independent and only mildly $T$ dependent, while
$M(T,H)$ varies strongly with $T$ and is nonlinear in $H$, the 2 contributions
are easily separated.  

\begin{figure}
\includegraphics[scale =0.35]{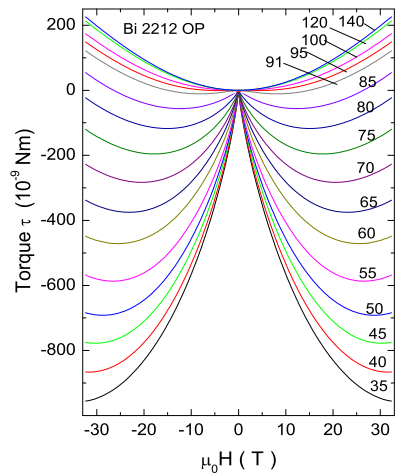}
\caption{  	\label{torque}  (color online) Curves of torque $\tau$ vs. $H$ in OP Bi 2212.  At the highest
$T$ (140 K), the magnetization is paramagnetic ($M = \chi_p H$), and 
$\tau\sim H^2$.  As $T$ decreases towards $T_c$ = 86.5 K, a negative 
diamagnetic contribution becomes apparent and grows rapidly to pull 
the torque negative.  Hystereses is large below 35 K.
}
\end{figure}

Figure \ref{torque} shows how $\tau$ varies with $H$ to 32 T in OP Bi 2212.  Above 120 K, 
only the paramagnetic term $\Delta\chi_p$ is visible.  Below 120 K, the diamagnetic term $M$
increases rapidly to pull the cantilever deflection to large negative values as $T$ decreases
below $T_c$ (86.5 K).  

\begin{figure}
\includegraphics[scale =0.5]{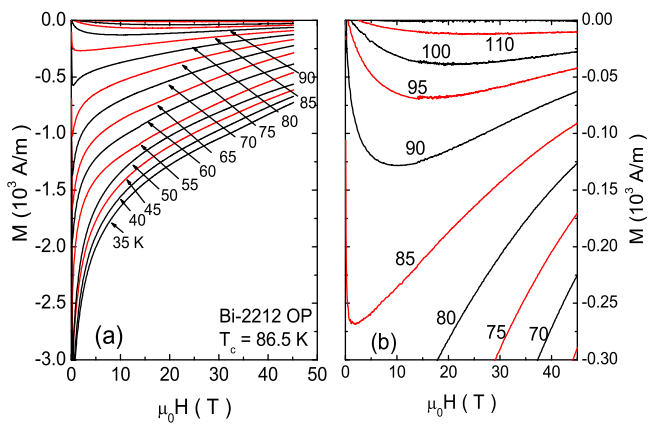}
\caption{  	\label{MH}  (color online) Magnetization curves $M$ vs. $H$ in OP Bi 2212 obtained from 
$\tau$ shown in Fig. \ref{torque}.  The right panel shows curves above 80 K in expanded scale.
At low $T$ (left panel), the field at which $M$ extrapolates to zero ($H_{c2}$) is estimated to be
150-200 T.  Note that as $T\rightarrow T_c^{-}$, $H_{c2}$ does not decrease below 45 T. 
}
\end{figure}

Dividing $\tau$ by $B_x$ and subtracting the term $\Delta\chi_p H$, 
we isolate $M(T,H)$ which is plotted in Fig. \ref{MH}.
Below 70 K, the $M$-$H$ curves closely resemble the Abrikosov profile familiar in low-$T_c$
superconductors; $M\sim \log H$ in the very large field interval $H_{c1}\ll H \ll H_{c2}$.  
As $T\rightarrow T_c^{-}$, however, a striking deviation from mean-field behavior
becomes apparent.  Slightly below $T_c$, the derivative $dM/dH$ in weak $H$ changes abruptly from 
positive to negative at a ``separatrix'' temperature $T_s$.  At $T_s$, $M(H)$ jumps 
abruptly at $H = 0^{+}$ to a finite value that is $H$-independent to fields 
of 5-7 T ($T_s$ is 2-3 K below $T_c$).  Above $T_c$, $M$ increases as the 
fractional power law $M\sim H^{1/\delta(T)}$, 
where the exponent $\delta(T)>1$ is anomalous and very strongly $T$ dependent.
In the interval where $\delta>1$ (between $T_c$ and 105 K), the system seems to exhibit
a fragile London rigidity which is easily destroyed in finite $H$.  The profile
$M$ vs. $H$ matches that of the Nernst profile $e_N$ vs. $H$ over a broad interval of $T$.
The features above $T_c$ are discussed in detail in Refs. \cite{LiEPL05} and \cite{LiM2S06}.  

Recently, the magnetization curves for a 2D large-$\kappa$ superconductor in the vicinity 
of its KT transition was calculated by Oganyesan, Huse and Sondhi~\cite{Oganesyan}.  For $T<T_{KT}$, the 
calculation reproduces several of the unusual features in Fig. \ref{MH} including the separatrix curve
and the change in sign of $dM/dH$ in weak $H$ on both sides of the separatrix.  However, 
above $T_{KT}$, the theory does not account for the anomalous exponent $\delta(T)$
discussed above ($M$ is always linear in $H$).

\begin{figure}
\includegraphics[scale =0.55]{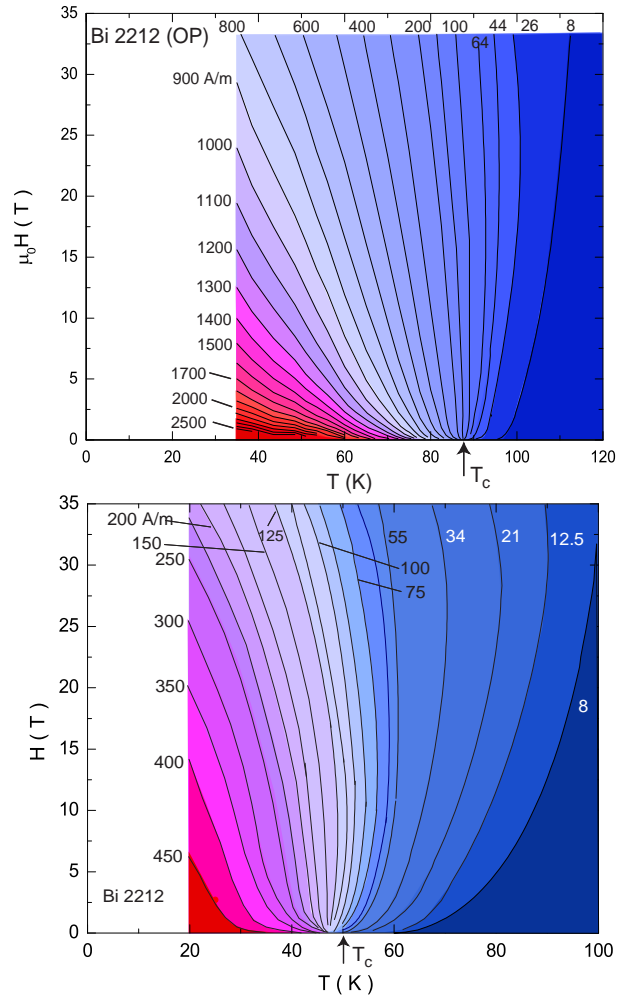}
\caption{  	\label{contour}
(color online) Contour plots of $|M(T,H)|$ in OP Bi 2212 ($T_c$ = 86.5 K, upper panel)
and in UD Bi 2212 ($T_c$ = 50 K, lower panel). In upper (lower) panel, the contour 
lines below $T_c$ are spaced 100 (50) A/m apart.  Above $T_c$ in both panels,
they are as indicated.  At the separatrix temperature $T_s$ (85 K and 47 K in upper and lower panels,
respectively), the contour is vertical up to 5 T. 
}
\end{figure}

As in the case of $e_N$, we may display $M(T,H)$ as a contour plot in the $T$-$H$ plane 
(Fig. \ref{contour}).  Above $\sim$100 K in the upper panel, 
the magnitude $|M|$ in OP Bi 2212 is small ($<$8 A/m;
deep blue region).  As $T$ decreases below $T_c$ to 35 K, $|M|$ rises to values $>$2,000
A/m (red region).  Let us recall that, in the MF (mean-field) transition, 
the contours above $T_c$ converge radially
to the point $(T,H) = (T_c,0)$, while below $T_c$ they are compressed into the $H_{c2}$ curve which
is a straight line terminating at $(T_c,0)$.  Here, the pattern is very different.  The contours are 
roughly parallel and vertical near $T_c$ except in very low $H$ where they converge
to $(T_c,0)$ non-analytically.  At $T_s$ (85 K), the contour is strictly vertical, reflecting the
constancy of $M$ vs. $H$ below $\sim$5 T.  The variation of the magnitude over
the whole $T$-$H$ region is also instructive.  In the MF transition, $|M|$ should drop
sharply to near-zero at the MF line $H_{c2}\sim (1-t)$, with $t=T/T_c$.  Instead, $|M|$ here
varies relatively slowly over the whole plane, retaining significant amplitude up to 120 K,
which corresponds to the vortex signal detected earlier in the Nernst experiment.
There is no evidence for a sharp boundary terminating at $T_c$ 
corresponding to the MF $H_{c2}(T)$ curve (extrapolation of the $M$-$H$ curves
in Fig. \ref{MH} yields values of $H_{c2}$ = 100-150 T even near $T_c$.
This anomalous behavior of $H_{c2}$ was pointed out long ago 
for the KT transition~\cite{Doniach}.  It may be generic to superconductors 
undergoing phase-disordering transitions.
By comparison, the contour features are even more strikingly anomalous in UD Bi 2212 (Fig. \ref{contour}, lower panel).
Above $T_c$ (50 K), $|M|$ is observable over a broader interval of $T$. 
The more pronounced curvature of the contour lines (relative to the OP sample) reflects
the larger temperature interval above $T_c$ where phase fluctuations exist.
The non-analytic behavior in weak fields around $(T_c,0)$ is also more evident.

\section{The depairing field and binding energy}\label{depairing}
The torque experiments allow $M$ vs. $H$ to be measured directly to fields as high as 45 T,
which is the scale of $H_{c2}(0)$ in UD, single-layer cuprates (but still quite a bit
smaller than in Bi 2212 as evident in Fig. \ref{MH}).  The $M$-$H$ curves in UD Bi 2201 
are reported in Ref.~\cite{LiM2S06}.

An important quantity may be derived by integrating the $M$-$H$ curve.
In BCS theory, the integral $\int_0^{H_{c2}} \; M(H)\;dH$ is the condensation energy 
of the superconducting state $E_c$.  This identity, based on thermodynamic
arguments, should be valid in a phase-disordered superconductor 
in the limit $T$ = 0.  The integration is especially accurate in the  
Bi-based cuprates and in UD LSCO where hystereses are negligible at large $H$ and low $T$.  
For single layer Bi 2201, we find (at 4 K) $E_c$ = 2,600 J/m$^3$ in an UD sample with $T_c$ = 12 K,
whereas $E_c$ = 5,600 J/m$^3$ in an OP sample with $T_c$ = 28 K. 
These values are significantly smaller than in OP bilayer Bi 2212 with $T_c$ = 86.5 K,
where we measure $E_c \sim 6\times 10^{5}$ J/m$^3$.  (For comparison, in Al, In, Pb and Nb, $E_c$ = 39,
341, 2,560 and 15,600 J/m$^3$, respectively.)

Assuming that the hole density $n_h$ = 0.15/Cu in OP Bi 2201 and 
and 0.22/Cu in OP Bi 2212, we calculate the condensation energy per hole $E_c/n_h$ 
to be 29 and 41 $\mu$eV, respectively.  The corresponding value
for Bi 2212 (2.1 meV per hole) is surprisingly large.  The large jump in $E_c/n_h$
between single and bilayer systems is not understood.

\section{Low-temperature vortex liquid}\label{lowT}
As one crosses the SC-dome boundary moving up in temperature at fixed $x$, the 
loss of phase coherence occurs via the spontaneous unbinding of vortex-antivortex pairs 
driven by the gain in entropy, in analogy with the 2D KT transition.  It is interesting to cross 
the boundary by decreasing $x$ below the critical value $x_c \sim$ 0.055
at very low $T$.  As $x$ approaches the Mott limit $x$ = 0, increased localization of the
Cooper pairs implies that local fluctuations in the pair density $\Delta n({\bf r})$ decreases.
Hence the conjugate variable, the phase $\theta$, fluctuates strongly.  At very low $T$,
this happens by the rapid motion of quantum vortices.  If the vortex solid is unstable,
long-range phase coherence is not possible even when $T\rightarrow 0$.

\begin{figure}
\includegraphics[scale =0.8]{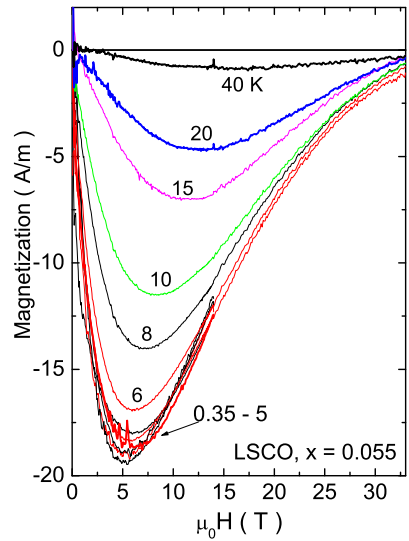}
\caption{  	\label{MlowT} (color online)
Magnetization curves $M$ vs. $H$ in LSCO ($x$ = 0.055) at selected $T$ below 40 K.
While the overall magnitude $|M|$ increases rapidly from 40 K to 5 K, it ceases
to change below 5 K.  The $M$-$H$ profile resembles that of the Nernst signal (Fig. \ref{Nernst}).
}
\end{figure}

To explore this interesting issue, we have extended the torque
experiment to below 1 K.  Below $\sim$6 K, a new 
contibution to the torque signal arises from the weakly anistropic paramagnetic
response of local moments (spin $s = \frac12$) given by
\begin{equation}
\Delta M_p = {n_s} \frac{\Delta g(T)}{2}\mu_B\tanh[\beta g(\theta)\mu_B{sH}],  
\;\; (\beta = 1/k_BT)
\label{para}
\end{equation}
with $\mu_B$ the Bohr magneton and $k_B$ Boltzmann's constant.  Here, $n_s$ is the 
density of the local moments and $\Delta g = g_c-g_{ab}$, where $g_c$ 
($g_{ab}$) is the g-factor measured with $\bf H||c$ 
($\bf H$ in the $ab$ plane).  In LSCO for $x<x_c$, we find that $\Delta g(T)$ is unobservable
until $T$ falls below $\sim$15 K.  

After this spin contribution is removed, we obtain the magnetization curves
associated with supercurrents in the CuO$_2$ layers shown in
Fig. \ref{MlowT} (in a crystal with $x$ = 0.055).  At 40 K, a weak diamagnetic 
signal with a profile that peaks near 5 T becomes evident.  The maximum 
value of $|M|$ increases as $T$ decreases
to 5 K.  However, as $T$ decreases to 0.35 K, no further change in $|M|$ is observed.  
We note that the curve bears a close resemblance to the tilted-hill profile of the Nernst signal
(Fig. \ref{Nernst}).
The diamagnetic signal extends to a field of $\sim$37 T which we
identify with the depairing field $H_{c2}(0)$.  We may contrast this behavior with that in
a crystal inside the SC dome (with $x$ = 0.060).  There, $M$ grows to very large 
values and exhibits hysteretic behavior when the vortices enter the solid state.

The low-$T$ magnetization reveals a sharp qualitative difference between samples
inside the SC dome and just outside.  In both cases, the depairing field is very large
(20-40 T), so that tightly bound pairs exist.  However, decreasing the temperature to 0 
has very different effects.  Inside the dome ($x>x_c$), the vortices 
enter the solid phase with an irreversibility field $\sim$8 T, and large hysteresis is observed.
If $x$ lies outside the dome, cooling has no observable effects on the $M$-$H$ curves below $\sim$6 K.
The vortices remain in the liquid state, and long-range phase coherence
is absent down to 0.5 K.  We interpret these results as evidence for the existence of a 
quantum vortex liquid in which fluctuations in $\theta$
prevent the establishment of long-range phase coherence even in the limit $T$ = 0.
This implies localization of the pairs.  The paramagnetic signal (Eq. \ref{para}) suggests
how this comes about.

At $x$ = 0.055, a small fraction $f_s$ of the holes enter a state that displays a paramagnetic 
magnetic signature consistent with nearly free local moments of 1 $\mu_B$ ($s=\frac12$) 
while the remainder remain Cooper-paired.  
As $x$ further decreases to 0.04 and 0.03, the fraction $f_s$ grows at 
the expense of the diamagnetic signal, suggesting a progressive conversion into the 
paramagnetic state.  By $x$ = 0.03, this magnetic state begins
to display magnetic hysteresis below 2 K suggestive of glassy behavior (the magnetic hysteresis
is easily distinguished from the hysteresis of the vortex solid seen only above $x_c$).  
Details will be published elsewhere.

\section{Spin gap, charge pairing and onset temperature}
As shown in Fig. \ref{phasediag}, the Nernst region occupies a large area that extends above the SC
dome into the pseudgogap state.  The electronic properties in
the Nernst region are highly unusual.  Although the in-plane resistivity $\rho_a$ is high and 
nominally $T$-linear, both the Nernst and diamagnetic signals increase 
steeply as $T\rightarrow T_c^{+}$.  Significant pair condensate amplitude exists in this region
but the high concentration of thermally generated mobile (anti)vortices reduces 
phase rigidity to very short length scales $\xi_{\phi}$ in zero $H$, so that the Meissner effect
is absent altogether.  

In discussions of the pseudogap, it seems important to recognize that the spin and 
charge degrees are affected differently.  As shown in Fig. \ref{phasediag}, the pseudogap 
temperature $T^*$ lies above $T_{onset}$.  Hence
when the pseudogap first appears (at 200-300 K) there is no evidence for Cooper pairing.
At such high $T$, the pseudogap is actually a spin gap.  The pseudogap was first 
inferred from the $T$ depedence of the relaxation rate $1/(T_1T)$ 
and the Knight shift $\Delta K_s$ in NMR experiments.  Cooper pairing involving
the charge degrees occurs only below $T_{onset}$, as evidenced by the steep rise of
the vortex-Nernst and diamagnetic signals.  This implies that the spin degrees sense the pairing instability
long before the charge degrees.  When the charges pair, vorticity and diamagnetism become
detectable, i.e. local ``superconductivity'' with very short $\xi_{\phi}$ appears.  
In this regard, it is significant that measurements of a ``pseudogap''
above $T_c$ by ARPES and tunneling (which probe the charge degrees) are 
confined to $T<$ 110 K in Bi 2212.  We suggest that these experiments are just detecting the 
gap of the superconducting condensate, but in its phase-disordered state.  

Recently, the important role of vortices in the UD region has been 
emphasized in several theories~\cite{Tesanovic,Weng,Honerkamp,Chen,Sachdev,Balents,Anderson06}.
An interesting theory for $T_{onset}$ has been proposed by Anderson~\cite{Anderson06}.
Hartree-Fock factorization of the $tJ$ Hamiltonian yields 2 self-energies
$\tilde{\Delta}_{\bf k}$ and $\tilde{\zeta}_{\bf k}$, which may be represented
in the space ($\cos k_x$, $\cos k_y$) as the orthogonal vectors~\cite{Zhang}
\be
\tilde{\Delta}_{\bf k}\sim (\cos k_x - \cos k_y), \quad\quad
\tilde{\zeta}_{\bf k}\sim (\cos k_x + \cos k_y).
\label{gap}
\ee  
A key feature is that the self energy $\tilde{\zeta}_{\bf k}$ shares the same 
``extended $s$-wave'' form as the qp kinetic energy.  At high temperatures ($T_{onset}<T<T^*$),
$\tilde{\zeta}_{\bf k}$ is unaligned with the kinetic energy, but
``locks'' to it at $T_{onset}$.  Below $T_{onset}$, the self-energy 
$\tilde{\Delta}_{\bf k}$ (the superconducting gap parameter),
continues to fluctuate in phase until $T_c$ where phase coherence becomes long-range.

Finally, we mention theories that propose that the Nernst signals (even those below $T_c$) 
are not produced by vortices at all, but by quasiparticles that occupy highly unusual electronic states --
either a charge density wave with a $d$-wave gap (DDW)~\cite{Dora}, or 
a band of uncondensed bipolaron bosons~\cite{Alexandrov}.  
In our view, these qp-based theories are not viable.  ``Fits'' to Nernst or magnetization data   
seem to be based on \emph{ad hoc}, unrealistic assumptions.  
The collective evidence, notably the adiabatic continuity from below to above $T_c$,
the strong correlation of the Nernst and diamagnetic signal with the 
SC dome, the accurate scaling between the Nernst and diamagnetic signals above $T_c$,
and the high-field suppression of both signals at $H_{c2}$ present a strong case for
a vortex-liquid origin (see Ref. \cite{WangPRB06}).

\vskip12pt
We have benefitted from discussions with P. W. Anderson, Z. Te\v{s}anovi\'c, S. A. Kivelson,
and J. C. Davis.  Research at Princeton was supported by the U.S. National Science Foundation (NSF)
under a MRSEC grant DMR 0213706.  Research at CRIEPI was supported by a 
Grant-in-Aid for Science provided by the Japan Society for 
the Promotion of Science.The high-field experiments were performed at the National
High Magnetic Field Laboratory, Tallahassee, which is supported by NSF, the U.S. 
Department of Energy and the State of Florida.


\begin{thebibliography}{00}

\bibitem{Schrieffer} See \emph{Theory of Superconductivity}, J. R. Schrieffer (Addison Wesley, 1964), ch. 8.


\bibitem{Anderson66} P. W. Anderson, Rev. Mod. Phys. {\bf 38}, 298-310 (1966).

\bibitem{KT} J. M. Kosterlitz and D. J. Thouless, J. Phys. C {\bf 6}, 1181 (1973).

\bibitem{Baskaran} G. Baskaran, Z. Zou and P. W. Anderson, Solid State Commun. {\bf 63}, 973 (1987).

\bibitem{Inui} S. Doniach and M. Inui, Phys. Rev. B {\bf 41}, 6668 (1990).

\bibitem{Uemura} Y. Uemura \emph{et al.}, Phys. Rev. Lett {\bf 62}, 2317 (1989); 
Y. Uemura \emph{et al.}, Nature {\bf 364}, 605 (1993).

\bibitem{Emery} V. J. Emery and S. A. Kivelson, Nature {\bf 374}, 434 (1995).

\bibitem{Corson} J. Corson \emph{et al.}, Nature {\bf 398}, 221 (1999).

\bibitem{Xu} Z. Xu \emph{et al.}, Nature {\bf 406}, 486 (2000).

\bibitem{WangPRB01} Yayu Wang \emph{et al.}, Phys. Rev. B {\bf 64}, 224519 (2001).

\bibitem{WangPRL02} Yayu Wang \emph{et al.}, Phys. Rev. Lett. {\bf 88}, 257003 (2002).

\bibitem{WangSci03} Yayu Wang \emph{et al.}, Science {\bf 299}, 86 (2003).

\bibitem{WangPRB06}  Yayu Wang, Lu Li and N. P. Ong, Phys. Rev. B {\bf 73}, 024510 (2006).  

\bibitem{WangPRL05} Yayu Wang \emph{et al.}, Phys. Rev. Lett.  {\bf 95}, 247002 (2005).

\bibitem{LiEPL05} Lu Li \emph{et al.}, Europhys. Lett. {\bf 72}, 451-457 (2005).

\bibitem{LiM2S06} Lu Li \emph{et al.}, \emph{Proceedings of
M$^2$S-HTSC-VIII, Dresden}, Physica C, \emph{in press}.

\bibitem{Oganesyan}  Vadim Oganesyan, David A. Huse and S. L. Sondhi, Phys. Rev. B {\bf 73}, 094503 (2006).

\bibitem{Doniach} S. Doniach and B. A. Huberman, Phys. Rev. Lett. {\bf 42}, 1160 (1979).

\bibitem{Tesanovic} Ashot Melikyan, Zlatko Tesanovic,
Phys. Rev. B {\bf 71}, 214511 (2005).

\bibitem{Weng} Z. C. Gu and Z. Y. Weng, Phys. Rev. B {\bf 72}, 104520 (2005).

\bibitem{Honerkamp} C. Honerkamp and P. A. Lee, Phys. Rev. Lett. {\bf 92}, 177002 (2004).

\bibitem{Chen} H. D. Chen, O. Vafek, A. Yazdani and S. C. Zhang, Phys. Rev. Lett. {\bf 93}, 187002 (2004).

\bibitem{Sachdev} S. Sachdev and E. Demler, Phys. Rev. B {\bf 69}, 144504 (2004).

\bibitem{Balents}  Leon Balents, Lorenz Bartosch, Anton Burkov, Subir Sachdev, and K. Sengupta
Phys. Rev. B {\bf 71}, 144508 (2005).


\bibitem{Anderson06} P. W. Anderson, Phys. Rev. Lett. {\bf 96}, 017001 (2006). 

\bibitem{Zhang} F. C. Zhang, C. Gros, T. M. Rice, and H. Shiba, J. Supercond. Sci. Tech. {\bf 1},
36 (1988), cond-mat/0311604.

\bibitem{Dora} B. Dora \emph{et al.}, Phys. Rev. B {\bf 68}, 241102(R) (2003).

\bibitem{Alexandrov} A. S. Alexandrov and V. N. Zavaritsky, Phys. Rev. Lett. 
{\bf 93}, 217002 (2004). 


\end{thebibliography}
\end{document}